%% file: sn-article.tex
\documentclass[a4paper, pdflatex, sn-mathphys-num]{sn-jnl}

\usepackage{graphicx}%
\usepackage{multirow}%
\usepackage{amsmath,amssymb,amsfonts}%
\usepackage{amsthm}%
\usepackage{mathrsfs}%
\DeclareFontFamily{U}{rsfs}{\skewchar\font127}
\DeclareFontShape{U}{rsfs}{m}{n}{
   <-6>    rsfs5
   <6-8>   rsfs7
   <8->    rsfs10
}{}
\usepackage[title]{appendix}%
\usepackage{xcolor}%
\usepackage{textcomp}%
\usepackage{manyfoot}%
\usepackage{booktabs}%
\usepackage{algorithm}%
\usepackage{algorithmicx}%
\usepackage{algpseudocode}%
\usepackage{listings}%
\usepackage{tikz}
\usetikzlibrary{calc}
\usepackage{float}
\usepackage{pgfplots}
\pgfplotsset{compat=1.18}
\usetikzlibrary{positioning, arrows.meta, decorations}

\usepackage[backend=biber, sorting=none, style=ieee]{biblatex}
\newcommand{\methodname}{Ours}

\theoremstyle{thmstyleone}%
\theoremstyle{thmstyletwo}%
\theoremstyle{thmstylethree}%

\title{Unsupervised Deep Learning for Inverse Problems in Computed Tomography}

\definecolor{mygreen}{RGB}{220,240,235}
\definecolor{mylightblue}{RGB}{204,222,229}
\definecolor{myorange}{RGB}{255,152,0}

\author[1,2,*]{Laura Hellwege}
\author[2]{Johann Christopher Engster}
\author[2]{Moritz Schaar}
\author[1,2]{Thorsten M. Buzug}
\author[2]{Maik Stille}
\affil[1]{Institute of Medical Engineering, University of Lübeck, 23562 Lübeck, Germany}
\affil[2]{Fraunhofer Research Institution for Individualized Medical Technology and Engineering (IMTE), 23562 Lübeck, Germany}
\affil[*]{l.hellwege@uni-luebeck.de}

\abstract{
Assume you encounter an inverse problem that shall be solved for a large number of data, but no ground-truth data is available. To emulate this, in this study we assume it is unknown how to solve the imaging problem of Computed Tomography. We introduce an unsupervised deep learning framework that leverages the inherent similarities between iterative reconstruction, Deep Image Prior (DIP), and unrolled optimization schemes. Our specific contribution is a training framework for amortized reconstruction: After training on a dataset without any image-domain ground truth, reconstruction of an unseen scan reduces to a single network forward pass.
We demonstrate the feasibility of reconstructing images from measurement data by pure network inference, without additional gradient steps for unseen samples.
Our method is evaluated on the two-dimensional 2DeteCT dataset. Within a controlled, geometry-matched benchmark, our reconstructions are competitive with, or better than, filtered back-projection, maximum-likelihood reconstruction, and a supervised network of identical architecture. Compared to a per-image DIP baseline, our method reaches similar quality while replacing the costly per-instance optimization with a single forward pass, yielding a speed-up of about four orders of magnitude.
This makes it a promising candidate for time-critical imaging applications. Future work will address multi-dataset adaptability, counter-measures against over-smoothing, advanced uncertainty quantification, and further medical-imaging inverse problems.
}


\begin{document}

\maketitle

\section{Introduction}

\subsection{Background}

We consider the example of Computed Tomography (CT) reconstruction as a typical inverse problem in imaging. The goal is to recover the original image from its projections.
Let $f: \mathbb{R}^2 \to \mathbb{R}$ be the continuous function that represents the imaging object's attenuation values. Assuming a circular measurement trajectory, the geometric forward projection (FP) in computed tomography can be expressed by the Radon transform
\begin{equation}
\mathrm{FP}(f) = \int_{-\infty}^{\infty} \int_{0}^{2\pi} f(x, y) \cdot \delta(x \cos \gamma + y \sin \gamma - s) \, d\gamma \, ds,
\label{eq:Radon}
\end{equation}
where \(\delta\) denotes the Dirac function, \(\gamma\) the projection angle and \(s\) the coordinate of the projection path through the object.
This continuous form can be transferred to a linear discrete forward problem. Let $f \in \mathbb{R}^N$ be the discretized, vectorized form of the image $f$ and let $p \in \mathbb{R}^M$ be its FP. Let the discrete version of the Radon transform from Eq.~\ref{eq:Radon} be represented by a system matrix $A \in \mathbb{R}^{M\times N}$, that attributes a contribution of each pixel $f_j, \, j = 1,...,N,$ to each projection value $p_i,\, i=1,...M$. Then, the system matrix defines the discrete FP  $A f = p$ as well as the inverse problem
\begin{equation}
\label{eq:inverseProblem}
    \text{Find} \hspace{4pt} f , \hspace{8pt}  \text{s.t.} \hspace{8pt} A f = p.
\end{equation}


\noindent Iterative reconstruction (IR) is a method that aims to progressively improve the reconstructed image $f$ while adhering to Eq.~\ref{eq:inverseProblem}. Figure~\ref{fig:IterativeScheme} visualizes the general procedure.
Let \(\hat{f}^k \in \mathbb{R}^N\) be the reconstructed tomogram at the iteration step \(k \in \{0, \ldots, K\}\).
In particular, let \(\hat{f}^0\) be the initial image, usually selected as a constant image with a very small positive value.
To achieve a stepwise approximation of the current image \(\hat{f}^k\) to the underlying true image \(f^\star\), a comparison is made between the measurement \(p\) itself and the FP of the current image \(\hat{f}^k\), i.e. $p^k = A \hat{f}^k$. Assuming a Gaussian distribution of the underlying projection data, the metric of discrepancy or loss $\mathcal{L}$ is chosen as the $\mathcal{L}^2$-norm
\begin{equation}
\label{eq:loss}
    \mathcal{L}(\hat{f}^k| p) = \| p - p^k\|^2 = \| p - A \hat{f}^k\|^2.
\end{equation}
Due to overall differentiability of Eq.~\ref{eq:loss}, the discrepancy $\mathcal{L}$ can be minimized by iterative gradient-based optimization, i.e.
\begin{equation}
\label{eq:optimization}
    \hat{f}^{k+1} = \hat{f}^{k} + \alpha \cdot \mathcal{D}(\nabla \hat{f}^k ),
\end{equation}
until a predefined stopping criterion is reached. Here, the step width $\alpha$ and the search direction $\mathcal{D}$ are chosen by the respective optimization method, usually in dependence on the gradient $\nabla \hat{f}^k$.
Regarding implementation, it is important to ensure that the matrix $A^T$ satisfies the adjointness condition
\begin{equation}
\label{eq:adjoint}
    \langle Af, p \rangle_{\mathbb{R}^M} = \langle f, A^Tp \rangle_{\mathbb{R}^N} \quad \forall f \in \mathbb{R}^N, \forall p \in \mathbb{R}^M,
\end{equation}
where $\langle \cdot, \cdot \rangle$ denotes the scalar product of the Hilbert spaces $\mathbb{R}^N$ and $\mathbb{R}^M$. This condition needs to be fulfilled for gradient-based optimization since the adjoint matrix $A^T$ is used as the domain transform from projection to image space - known as back projection (BP). When computing the image gradient from the projection differences, adjointness of $A^T$ guarantees appropriate scaling of the gradient values~\cite{bertero2021}.

\begin{figure}[t!]
\include{gradient_based_reco_scheme}
\vspace{-1cm}
\caption{Gradient-based image reconstruction scheme. The colored boxes indicate the domain of the respective data (
        \begin{tikzpicture}[scale=0.3]
            \fill[orange!20!] (-0.25,-0.25) rectangle (0.5,0.5);
        \end{tikzpicture} projection domain,
        \begin{tikzpicture}[scale=0.3]
            \fill[mylightblue]  (-0.25,-0.25) rectangle (0.5,0.5);
        \end{tikzpicture} image domain).
        The boxes
        \begin{tikzpicture}[scale=0.3]
            \fill[gray!20!]  (-0.25,-0.25) rectangle (0.5,0.5);
        \end{tikzpicture} indicate the domain transform by FP ($A$) or BP ($A^T$). The steps to iteratively adapt the current image $\hat{f}^k$ with the current gradient (\textcolor{orange}{$\nabla$}) are indicated by arrows. The loss $\mathcal{L}$ is defined by Eq.~\ref{eq:loss}.
        }
\label{fig:IterativeScheme}
\end{figure}

\subsection{Related Work}
Deep learning (DL) has emerged as a transformative technology in various fields, including medical imaging. In the context of CT reconstruction, different methods have been developed to enhance image quality through supervised and unsupervised training techniques.

Many approaches learn the reconstruction map from paired data: Image-domain post-processing networks improve filtered back-projection (FBP) outputs~\cite{Jin2017}, sinogram-domain networks learn the filtering step~\cite{He2018}, FBP-inspired end-to-end networks map sinograms directly to images~\cite{Li2019}, dual-domain networks couple projection- and image-domain units~\cite{Ge2020}, and learned primal--dual schemes embed the forward operator into an unrolled optimization trained against ground truth~\cite{Adler2018}. While effective, all of these require paired ground-truth images, which are unavailable in typical medical settings.

To avoid the necessity of ground-truth , per-instance unsupervised methods optimize the network weights individually for each image. Deep Image Prior (DIP) exploits the implicit regularization of a convolutional architecture in this way~\cite{Ulyanov2017,Zhang2020, Li2020}, and Deep Radon Prior~\cite{Xu2024} follows a similar approach, combining neural networks with traditional iterative reconstruction and thus optimizing per dataset. 
Self-supervised strategies instead exploit measurement statistics or physics: They split the projections to build a self-supervised target~\cite{Hendriksen2020} and or partition the measured data per subject into disjoint sets for training and data-consistency \cite{Yaman2020}. This partitioning resembles cross-validation and is performed individually per subject; it is therefore not classical amortized inference across a dataset.
Physics-informed and unrolled networks such as MoDL \cite{Aggarwal2019}, Variational Networks \cite{Hammernik2018} and E2E-VarNet \cite{Sriram2020} embed the forward operator into repeated update blocks but are predominantly supervised, and Equivariant Imaging\cite{Chen2021}, though mathematically principled, relies on an equivariance condition with multiple distinct imaging operators that is unavailable in our fixed-geometry, single-operator setting.

Our framework differs from all of the above primarily in how the trained network
is used at inference. After unsupervised training with a projection-domain
$\mathcal{L}_2$ loss on many datasets without image-domain ground truth,
reconstruction of any new, unseen scan reduces to a single forward pass with no
per-instance optimization. The network operates by design as a learned
image-domain filter of the back-projected (BP) input for amortized single-pass inference over a dataset.

\subsection{Scope of this work}
Our main contribution is an unsupervised training framework for image reconstruction, treating it as an inverse problem.
We leverage the similarities of traditional network training with amortized inference, DIP and unrolled optimization schemes.
Our focus is the feasibility of training a neural network to solve an inverse problem using only the forward model of the imaging system, without relying on any ground-truth images.
By design, the network acts as an image-domain filter of the BP input. The filters behavior is learned for the given training dataset. 
To demonstrate the efficacy of our method we choose two experiments that learn the accurate filtering of blurred images from simple backprojection for two datasets:
\begin{enumerate}
    \item[\textbf{A}] Forward projections (Eq.~\ref{eq:Radon}) of provided FBP CT slices acquired from high-precision data. The image evaluation can then be performed with said FBP slices. This is a controlled benchmark with perfect forward-model knowledge: Because the same linear operator generates and inverts the data, the setup has the characteristics of an ``inverse crime'' and should be read as a best-case, methodological feasibility test. By design, the FBP slices are not a true ground-truth object information, but they serve as a the reference ground truth for quantitative evaluation in this case.
    \item[\textbf{B}] Real measurements of low-quality projection data. The image quality is assessed with no ground truth available. Model knowledge is assumed only by given geometry. Inherent mismatch towards underlying beam-hardening physics is tolerated to compare reconstruction quality against high-precision image data.

\end{enumerate}

\section{Methods}\label{sec:methods}

\subsection{Forward and backward projection layers}
\label{subsec:Parallelproj}
\noindent A critical point of novel deep learning approaches in CT is the incorporation of FP and BP operators into the deep learning framework.
These operators enable forward and backward pass computations from the image domain to the projection domain and vice versa.
The deep learning framework of this work is implemented in PyTorch\cite{Paszke2019}.
We employ the projector operator definitions of the open-source library Parallelproj~\cite{Schramm2023-sj}, which was originally designed for positron emission tomography projections utilizing PyTorch compatible tensors.
The projection geometry can be defined manually to emulate a circular CT projection trajectory by explicitly defining the respective start and end points of the projection rays.
Hence, a projection operator and its adjoint can be incorporated in custom FP and BP network layers.
These layers now take advantage of PyTorch's autograd function, such that no manual gradient computations are necessary. The FP and BP of Parallelproj are defined mathematically adjoint (Eq.~\ref{eq:adjoint}), such that iterative computations of gradients are numerically consistent.

\begin{figure}[t!]
\include{training_scheme}
\vspace{-1cm}
\caption{Proposed unsupervised training scheme. The colored boxes indicate the domain of the respective data (
        \begin{tikzpicture}[scale=0.3]
            \fill[orange!20!] (-0.2,-0.2) rectangle (0.8,0.8);
        \end{tikzpicture} projection domain,
        \begin{tikzpicture}[scale=0.3]
            \fill[mylightblue]  (-0.2,-0.2) rectangle (0.8,0.8);
        \end{tikzpicture} image domain).
        The boxes
        \begin{tikzpicture}[scale=0.3]
            \fill[gray!20!]  (-0.2,-0.2) rectangle (0.8,0.8);
        \end{tikzpicture} indicate the domain transform by FP ($A$) or BP ($A^T$). The steps to iteratively adapt the UNet++ weights are indicated by arrows. Symbols follow the main-text notation ($\tilde{f}$: BP input, $\hat{f}$: Network prediction, $\hat{p}$: Predicted projection).
        }
\label{fig:TrainingScheme}
\end{figure}

\subsection{Unsupervised training scheme}

Figure~\ref{fig:TrainingScheme} illustrates our proposed unsupervised training approach. As network input, we only assume knowledge of the projection data $p$, which is transformed to the image domain by a simple BP (Radon inversion) layer $A^T$.
The resulting blurred image $\tilde{f}$ is fed into a deep convolutional network, namely the U-Net++\cite{Zhou2018} with the {ResNeXt-101} (32 $\times$ 16d)~\cite{Xie2017} encoder.
U-Net++ extends the classical U-Net~\cite{Ronneberger2015unet} by nested and dense skip connections: Intermediate decoder nodes are connected through a series of convolutional blocks along re-designed skip pathways, which reduces the semantic gap between encoder and decoder feature maps. The ResNeXt-101 encoder aggregates a set of parallel transformations of the same topology (its ``cardinality''), here $32$ groups with a bottleneck width of $16$, which increases representational capacity at a controlled parameter budget. We emphasize that the architecture is interchangeable; an architecture-swap ablation uses a smaller backbone to confirm that the framework does not depend on this particular choice.
It is important to notice, that training is employed batch-wise on the whole training dataset, making it different from previously unrolled optimization schemes.
Regarding the implementation, the Segmentation Models PyTorch~\cite{Iakubovskii2019} library, which enables the training of the regression problem, was employed.\\
The network's output is the desired high-quality image reconstruction $\hat{f}$.
To enable unsupervised training, the forward pass through the network is completed with the FP layer $A$ that is applied on the image $\hat{f}$.
Thus, the reconstructed image $\hat{f}$ is represented in the projection data domain as $\hat{p} = A\hat{f}$.
Hence, $\mathcal{L}_2$-loss calculations can be performed in the projection domain without the need for an image-domain ground truth.
The backward pass propagates the calculated scalar loss value through the predicted projection tensor $\hat{p}$, the BP layer $A^T$ and the predicted image tensor $\hat{f}$ into the network layers.
For training, the Adaptive Momentum optimizer (ADAM) was used with a learning rate of $10^{-4}$, first- and second-moment decay rates $\beta_1 = 0.9$ and $\beta_2 = 0.999$, and $\epsilon = 10^{-8}$. No weight decay was applied ($\lambda = 0$), and the learning rate was kept constant throughout training (no scheduling), on an NVIDIA RTX A6000 GPU with 48 GB of VRAM. No augmentations were applied to maintain physical and geometrical consistency. Omitting augmentation is a deliberate trade-off that preserves the exact geometric relationship between the input $\tilde{f}$ and the projection data $p$ required by the projection-domain loss, but it may limit generalization compared to augmented training.
We utilized a batch size of 8 training samples due to VRAM limitations. During training, the currently best performing network with respect to the validation data was stored, and training was performed until the validation metrics did not decline for 100 epochs. In practice, convergence was reached after 330-480 epochs amounting to a runtime of approximately 45-67~h per training on the hardware above.

\noindent
Comparing Figure~\ref{fig:IterativeScheme} and Figure~\ref{fig:TrainingScheme}, the current training prediction $\hat{f}$ coincides with the gradient-based IR current image $\hat{f}^k$.
The IR loss and the training loss are computed with the same formula from Eq.~\ref{eq:loss}. This similarity was previously exploited for reconstruction of separated datasets in learned IR methods.
Instead of altering the image itself, in network training, the gradient only affects the network's weights.
In contrast to unrolled optimization and DIP methods, when training is performed with many datasets, the network should be able to generalize the filter step.
Hence, the main difference to learned IR methods is also the main advantage of our approach:
Once trained, the network computes the high-quality image from the blurry initial image $\tilde{f}$ in a single forward pass.
Compared to IR or learned IR, this eliminates the need for computationally extensive gradient calculations or auto-differentiation steps after training completely.

\subsection{Data}
\label{subsec:data}

For validation of the proposed unsupervised training approach, the publicly available 2DeteCT\cite{Kiss2023} dataset is employed. It includes over 5,000 CT reconstructed 2D FBP slices and their respective 2D projection data. The projections of each slice were acquired with three varying radiation parameter modes.
We considered two of the three given modes in this work:
High precision mode (90 kVp, Thoraeus filter of Sn=0.1 mm, Cu=0.2 mm, Al=0.5 mm) and beam-hardening affected mode (60 kVp, no filtration).

We employ 2800 of the 5000 provided slices and respective projection data, since a slight difference in geometry has been reported for the other measurements.
This poses a critical difference for the reconstruction network since the FP and BP layers, as of now, are designed as fixed.
To keep validation times during training short, we choose 100 projection datasets for validation. 2400 projection datasets are used for training. Approximately 10\% of the available data (300 test datasets) are used for evaluation. The train/validation/test split is patient-disjoint: Slices originating from the same physical sample are assigned entirely to one split only, so that no sample contributes to both training and evaluation. This rules out slice-level leakage between splits.

\subsection{Baselines}
\label{subsec:baselines}
To compare our approach to conventional methods, we employ FBP with a Hann filter (cut-off at 80\% of normalized maximal frequency) and gradient-based maximum likelihood (ML) reconstruction.
The FBP is calculated with our in-house software based on the \textit{Reconstruction Toolkit}\cite{Rit2014}.
The ML optimization is performed with \textit{SciPy}'s\cite{2020SciPy-NMeth} L-BFGS-B implementation. The ML baseline is run until convergence (norm of gradient $\leq 10^{-8}$) or with a maximal number of 400 iterations. 
A supervised deep learning approach with the same architecture is trained with the $\mathcal{L}_2$ loss with respect to the reference FBP images. This can be interpreted as the best-case deep learning approach for Experiment A.
As an additional comparison, we include a DIP baseline. Since DIP is optimized per image, we evaluate it on 10 random representative test slices only. We frame this comparison as illustrative since $n=10$ does not support robust quality statistics. 

\subsection{Ablation study setup}
We conduct an ablation study in three properties. (1) A dataset-size sweep with 300, 1200 and 2400 training slices to characterize data efficiency with (2) an architecture-swap using the smaller encoder backbone ResNet-18~\cite{He2015Resnet18} to demonstrate that the framework is not tied to a specific network and (3) a reduction of resolution from originally $1024\times1024$ to $256\times256$ pixels.

\subsection{Statistical analysis}
Quantitative comparisons across the 300 test slices use paired Wilcoxon-rank tests. To control the family-wise error rate over multiple comparisons, we apply the Holm--Bonferroni correction and report corrected $p$-values. We additionally report effect sizes (rank-biserial correlation) and bootstrap 95\% confidence intervals (CI). 

\subsection{Image quality assessment}
Quantitative image quality was assessed using the Noise Power Spectrum (NPS) and the Modulation Transfer Function (MTF).
The NPS characterizes image noise as a function of spatial frequency, computed via radially-averaged 2D Fourier analysis of uniform image regions. The MTF quantifies spatial resolution using the slanted-edge method: High-contrast edges were automatically detected using the Canny algorithm ($\sigma = 2.0$), and for each edge, the Edge Spread Function was extracted with $4\times$ subpixel oversampling. The Line Spread Function was obtained by differentiation of the ESF, and the MTF computed as the normalized magnitude of its Fourier transform.
Gaussian and exponential models were fitted to extract MTF$_{50}$ and MTF$_{10}$ values.
Because the MTF is estimated from real detected edges (multi-edge estimation), the reported values should be interpreted only relatively between methods, not as an absolute, resolution-invariant system metric. Analysis was performed on 10\% of the test data slices with approximately 20 edges per slice.

\section{Results}\label{sec:result}
\subsection{Experiment A: Reconstruction of FP data}

Experiment A was evaluated in multiple stages. Since this experiment was conducted with forward projected data for which the exact geometry was known for both forward and backward projection, we can assume perfect alignment of projected and reconstructed data with our reference.

\begin{figure}[b!]
    \centering
    \sffamily
    \resizebox{\linewidth}{!}{%
    \begin{tikzpicture}
        \node[inner sep=0pt] (img1) at (0, 0){\includegraphics[width=13cm]{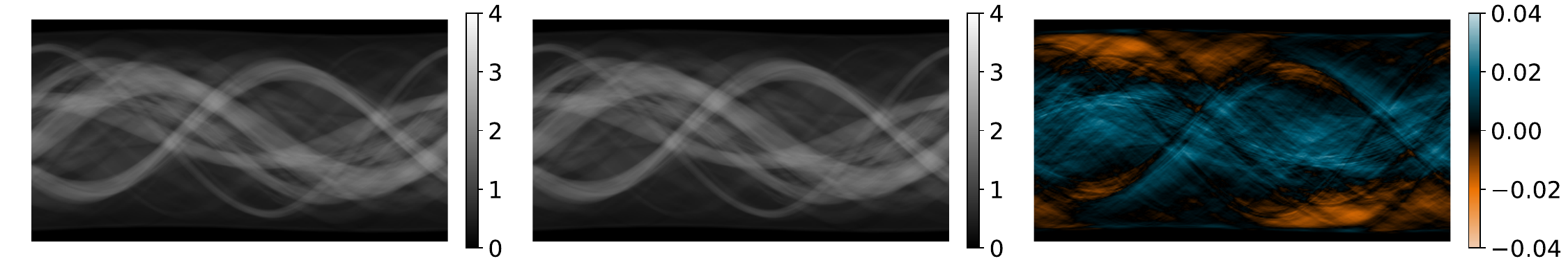}};
        \node at ([yshift=0.1cm, xshift=1.85cm]img1.north west) {\small (a) \methodname};
        \node at ([yshift=0.1cm, xshift=6.1cm]img1.north west) {\small (b) GT};
        \node at ([yshift=0.1cm, xshift=10.3cm]img1.north west) {\small (c) \methodname\ - GT};
        \node[xshift=-4.5cm, yshift=-1.1cm]{\footnotesize Gantry Angle};
        \node[rotate=90, xshift=+0.0cm, yshift=+6.5cm]{\footnotesize Detector index};
    \end{tikzpicture}
    }
    \vspace{-0.5cm}
    \caption{Exemplary visualization of (a) predicted projection data, (b) reference projection data, and (c) difference. Projection values are unit-less.}
    \label{fig:A-proj}
\end{figure}

\begin{figure}[ht!]
    \centering
    \begin{tikzpicture}
        \node[](img){\includegraphics[width=7cm]{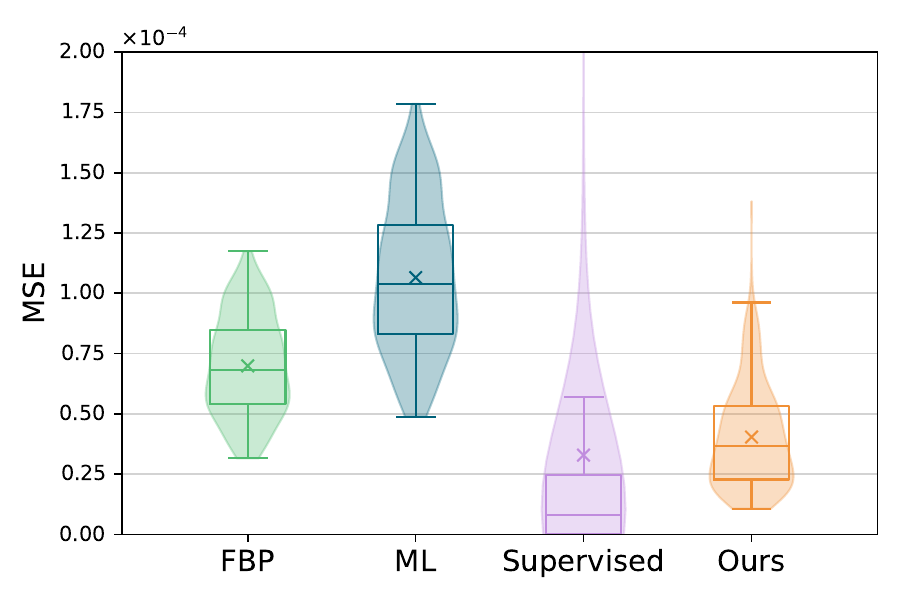}};
    \end{tikzpicture}
    \caption{MSE distribution of 300 projection predictions from the test dataset with respect to reference projection data. The crosses indicate the mean value for each method. Values in the projection domain are unit-less.}
    \label{fig:A-proj_quantitative}
\end{figure}

First, the general behavior of our method concerning projection domain loss minimization is investigated.
Figure~\ref{fig:A-proj} shows a qualitative example of the projection data network prediction versus the reference data. The exemplary difference image in Figure~\ref{fig:A-proj}(c) as well as the quantitative mean-squared-error (MSE) evaluation in Figure~\ref{fig:A-proj_quantitative} show that our method indeed predicts projection data that is highly similar to the reference projection data. Slight overestimation can be observed in the center of the sinogram and underestimation in its edge regions.

\begin{figure}[ht!]
    \centering
    \includegraphics[width=\linewidth]{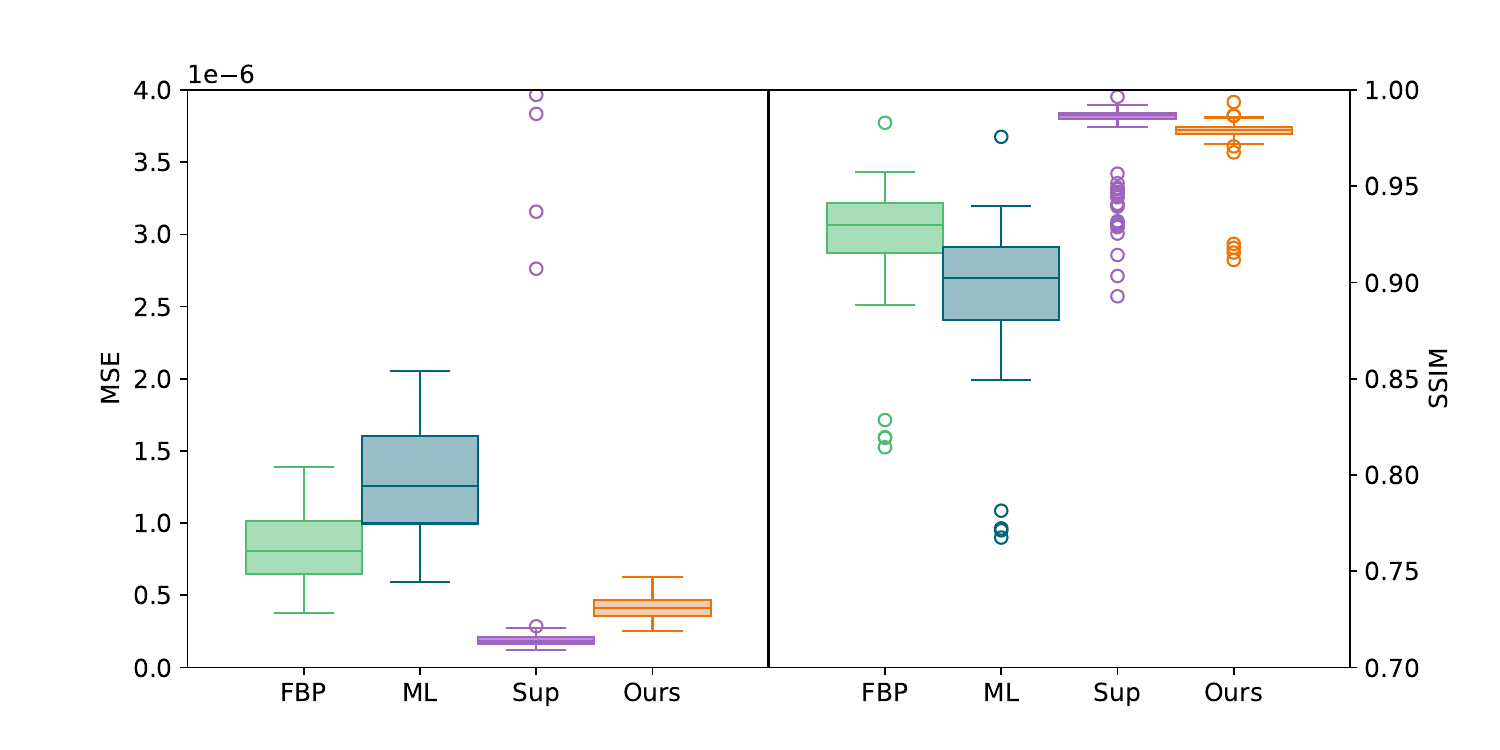}
    \vspace{-0.75cm}
    \caption{Quantitative results in image domain for the Radon inversion task on 300 test images. Deviations are calculated with respect to the reference image. Sup denotes the supervised method. [MSE] = $1/\mathrm{cm}^2$. 
    }
    \label{fig:A-quantitative}
\end{figure}

Second, the quantitative deviation of the reconstructed images to the reference images was evaluated.
The results for MSE and structural similarity (SSIM) are presented in Figure~\ref{fig:A-quantitative}.
It could be observed that the images reconstructed by our unsupervised network show lower MSE and higher SSIM values than those obtained with FBP/ML. The supervised method outperforms our approach, while producing noticeable high-deviation outliers in projection and image domain. These outliers can be attributed to missing image borders in the network predictions.
The relative trend of the distributions in Figure~\ref{fig:A-proj_quantitative} and Figure~\ref{fig:A-quantitative} is consistent across domains and metrics.

\begin{figure}
    \centering
    \includegraphics[width=5cm]{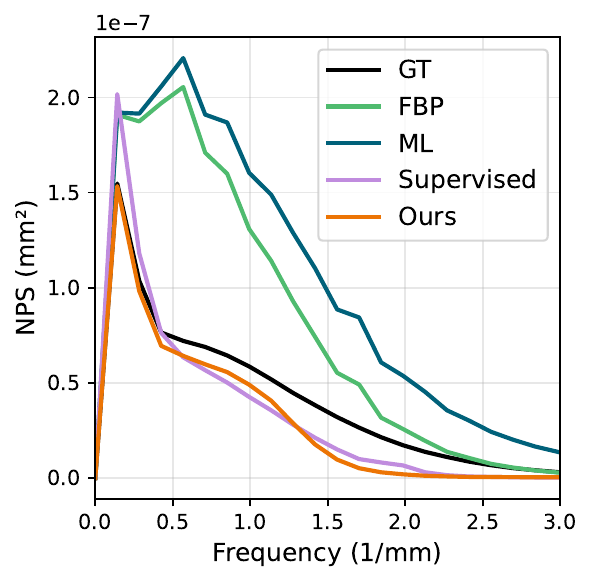}
    \includegraphics[width=5cm, trim=0cm 0cm 0cm 0cm, clip]{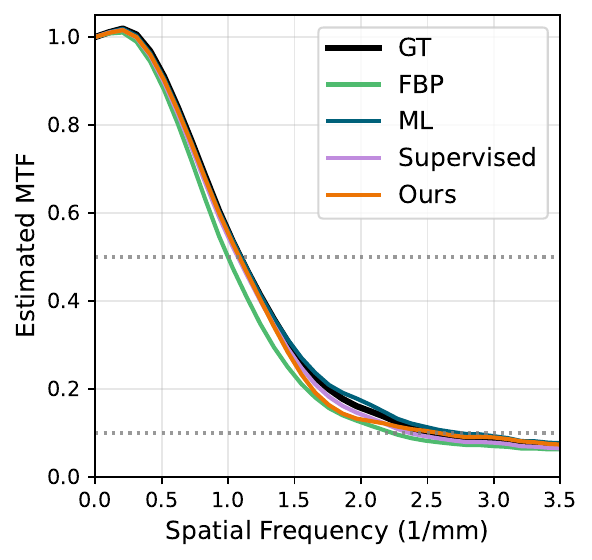}
    \caption{Radially-averaged NPS and estimated MTF for different methods. Horizontal lines at 0.5 and 0.1 indicate intersection points for MTF50 and MTF10 numbers.}
    \label{fig:A-NPSandMTF}
\end{figure}

Our unsupervised DL approach substantially outperforms both unsupervised baselines (FBP, ML) in terms of image MSE and SSIM ($p_{\mathrm{corr}} < 10^{-49}$, $r = 1.00$ in all four comparisons). Compared to the supervised method, however, our approach ranks second: The supervised approach achieves both a lower MSE  ($p_{\mathrm{corr}} < 10^{-28}$, $r = 0.74$, 95\% CI $[0.63, 0.84]$) and a higher SSIM ($p_{\mathrm{corr}} < 10^{-28}$, $r = 0.74$, 95\% CI $[0.63, 0.84]$). Nonetheless, the unsupervised DL approach attains competitive reconstruction quality without paired ground-truth supervision, closing much of the gap to the fully supervised upper bound while clearly surpassing all unsupervised baselines. 
The low MSE of the FBP baseline combined with its poor SSIM underscores the well-known limitation of pixel-wise metrics. While SSIM balances tractability and perceptual relevance, it captures neither fine diagnostic structures nor potential reconstruction artifacts reliably. We therefore complement it with NPS and MTF analysis in Figure~\ref{fig:A-NPSandMTF} to assess noise and spatial-resolution characteristics. We first assess the noise characteristics via the integrated NPS, where lower values indicate less residual noise. Our method achieves the lowest overall noise ($9.44\times10^{-8}$~mm$^2$), even below the reference ($1.27\times10^{-7}$~mm$^2$), followed by the supervised baseline ($1.05\times10^{-7}$~mm$^2$), whereas FBP ($2.49\times10^{-7}$~mm$^2$) and the model-based reconstruction ($3.22\times10^{-7}$~mm$^2$) remain substantially noisier.\\
Regarding spatial resolution, after per-slice outlier exclusion all methods
yield comparable MTF50 values around $1.0$--$1.1$~1/mm, with our method
(MTF50 $=1.08$~1/mm) matching the reference ($1.09$~1/mm) most closely.
The high-frequency behaviour (MTF10) reveals qualitative differences:
Our reconstruction ($2.59$~1/mm) again closely reproduces the reference
($2.53$~1/mm), while FBP ($2.24$~1/mm) and the supervised baseline
($2.37$~1/mm) both fall below the reference, indicating a loss of fine
detail, and ML rises to $2.74$~1/mm, suggesting a mild tendency towards edge
over-enhancement. Crucially, our method attains the lowest noise level while
preserving reference-like resolution at both the MTF50 and MTF10 levels,
demonstrating that its noise reduction does not come at the cost of spatial
detail.


\begin{figure}[ht!]
    \centering
    \sffamily
    \resizebox{\textwidth}{!}{%
    \begin{tikzpicture}
        \node[inner sep=0pt] (img) {\includegraphics[width=17.6cm]{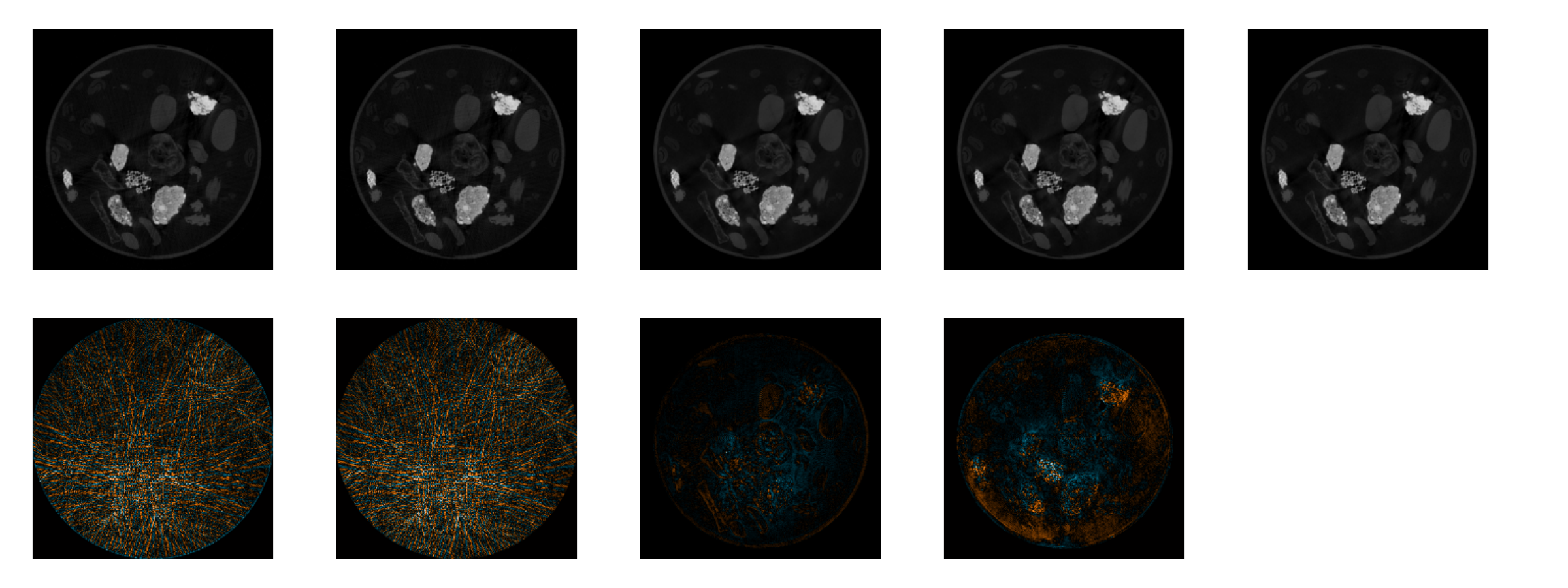}};
        \node at ([yshift=-0.25cm, xshift=-0.1cm]img.north west) {(a)};
        \node at ([yshift=-0.145cm, xshift=1.75cm]img.north west) {FBP};
        \node at ([yshift=-0.145cm, xshift=5.15cm]img.north west) {ML};
        \node at ([yshift=-0.155cm, xshift=8.55cm]img.north west) {Supervised};
        \node at ([yshift=-0.145cm, xshift=11.95cm]img.north west) { \methodname};
        \node at ([yshift=-0.145cm, xshift=15.35cm]img.north west) {GT};
        \node at ([yshift=-3.4cm, xshift=1.75cm]img.north west) {FBP - GT};
        \node at ([yshift=-3.4cm, xshift=5.15cm]img.north west) { ML - GT};
        \node at ([yshift=-3.41cm, xshift=8.55cm]img.north west) {Supervised - GT};
        \node at ([yshift=-3.4cm, xshift=11.95cm]img.north west) {\methodname\ - GT};
        \draw[orange, thick] (5.85,0.60) rectangle (6.95,1.7);
        \node at (5.5,-1.6) {
            \begin{axis}[
                anchor=north west, colorbar,
                colormap={custom_shift}{
                rgb255(0cm)=(240,205,179); rgb255(1cm)=(236,116,4);
                rgb255(2cm)=(0,0,0); rgb255(3cm)=(0,97,122);
                rgb255(4cm)=(198,220,226)},
                colorbar style={ width=0.5cm, height=2.7cm,
                    ytick={0,1,2,3,4},
                    yticklabels={$-0.004 \frac{1}{\mathrm{cm}}$, $+0.004 \frac{1}{\mathrm{cm}}$},},
                axis y line=none, axis x line=none,
                xmin=0, xmax=1, ymin=0, ymax=1,
            ]\end{axis}
        };
    \end{tikzpicture}
    }
\resizebox{\textwidth}{!}{%
 \begin{tikzpicture}
        \node[inner sep=0pt] (img) {\includegraphics[width=17.6cm]{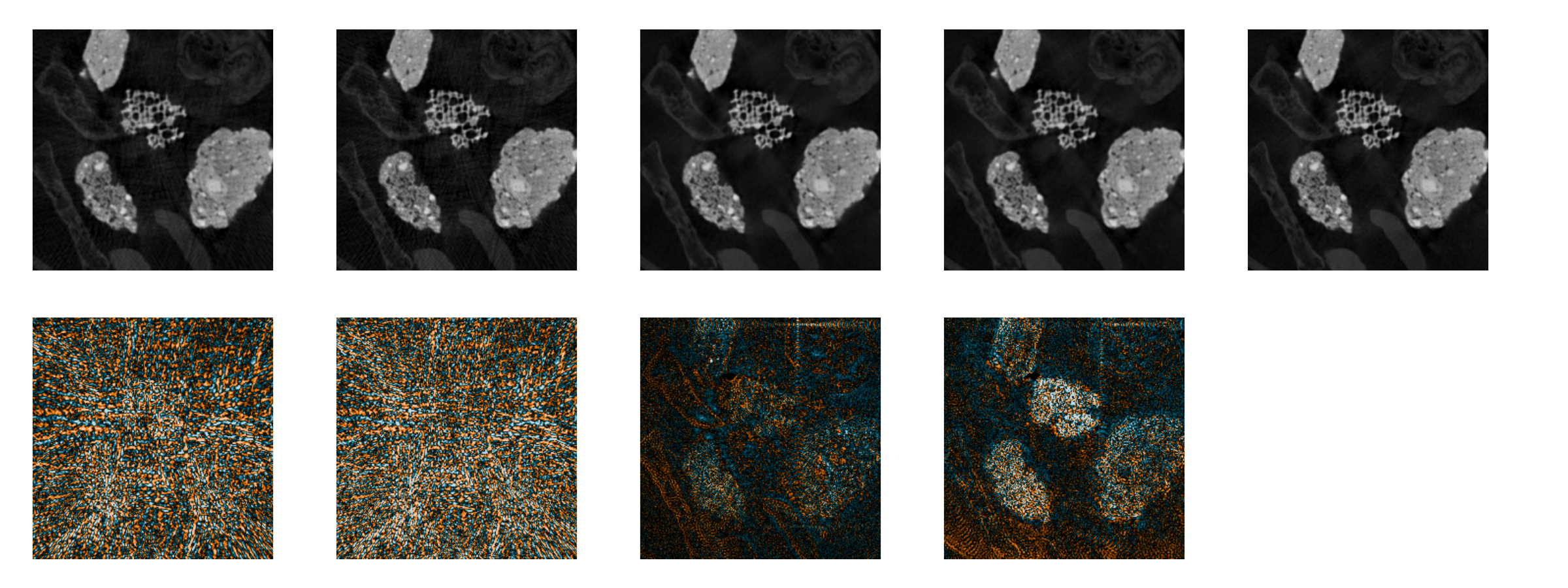}};
        \node at ([yshift=-0.25cm, xshift=-0.1cm]img.north west) {(b)};
        \draw[orange, thick] (5.19,0.27) rectangle (7.91,2.98);
        \draw[orange, thick] (5.17,0.25) rectangle (7.93,3.00);
        \node at ([yshift=-0.145cm, xshift=1.75cm]img.north west) {FBP};
        \node at ([yshift=-0.145cm, xshift=5.15cm]img.north west) {ML};
        \node at ([yshift=-0.155cm, xshift=8.55cm]img.north west) {Supervised};
        \node at ([yshift=-0.145cm, xshift=11.95cm]img.north west) { \methodname};
        \node at ([yshift=-0.145cm, xshift=15.35cm]img.north west) {GT};
        \node at ([yshift=-3.4cm, xshift=1.75cm]img.north west) {FBP - GT};
        \node at ([yshift=-3.4cm, xshift=5.15cm]img.north west) { ML - GT};
        \node at ([yshift=-3.41cm, xshift=8.55cm]img.north west) {Supervised - GT};
        \node at ([yshift=-3.4cm, xshift=11.95cm]img.north west) {\methodname\ - GT};
        \node at (5.5,-1.6) {
            \begin{axis}[
                anchor=north west, colorbar,
                colormap={custom_shift}{
                rgb255(0cm)=(240,205,179); rgb255(1cm)=(236,116,4);
                rgb255(2cm)=(0,0,0); rgb255(3cm)=(0,97,122);
                rgb255(4cm)=(198,220,226)},
                colorbar style={ width=0.5cm, height=2.7cm,
                    ytick={0,1,2,3,4},
                    yticklabels={$-0.003 \frac{1}{\mathrm{cm}}$, $+0.003 \frac{1}{\mathrm{cm}}$},},
                axis y line=none, axis x line=none,
                xmin=0, xmax=1, ymin=0, ymax=1,
            ]\end{axis}
        };
    \end{tikzpicture}
    }
     \caption{(a) Exemplary reconstructed images in L: 0.03\(\frac{1}{\mathrm{cm}}\),  W:0.06 \(\frac{1}{\mathrm{cm}}\) and (b) ROIs in L: 0.05\(\frac{1}{\mathrm{cm}}\),  W:0.1 \(\frac{1}{\mathrm{cm}}\) with difference images to reference ground-truth (GT) for different reconstruction methods. The image was chosen as one of the worst MSE values (MSE = \(6.8 \cdot 10^{-7} \, \frac{1}{\mathrm{cm}^2}\)) between our method's result and the GT image.}
    \label{fig:A-qualitative}
\end{figure}

Third, the reconstructed images were evaluated qualitatively. In Figure~\ref{fig:A-qualitative}, a slice of high deviation to the reference image was evaluated. High image quality can be observed in all reconstructions. The difference images reveal that while FBP and ML exhibit a typical noise level all over the reconstructed image, the predicted images for the supervised and our unsupervised approach mostly show differences in higher density regions.

\subsection{Experiment B: Reconstruction of real data}

\begin{figure}[]
\sffamily
    \centering
    \resizebox{\textwidth}{!}{
    \begin{tikzpicture}
        \node[](baselines){\includegraphics[width=8cm]{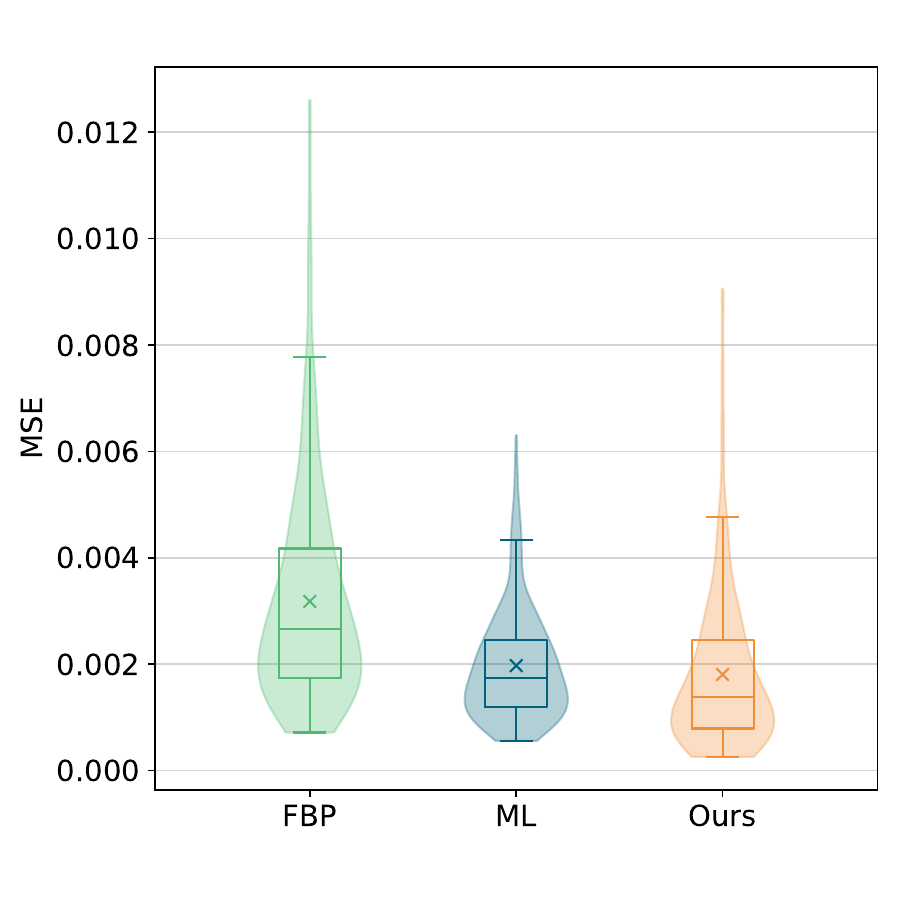}};
        \node[yshift=2.4cm, xshift=-1.5cm, align=center]{$\bar{t}_\text{FBP}$ \\ $\approx$ 1.7 s};
        \node[yshift=2.4cm, xshift=+0.5cm, align=center]{$\bar{t}_\text{ML}$ \\ $\approx$ 63.4 s};
        \node[yshift=2.4cm, xshift=+2.5cm, align=center]{$\bar{t}_\text{Ours}$ \\ $\approx$ 0.4 s};
        \node[above=of baselines, yshift=-1.75cm, xshift=-3.25cm]{(a)};

        \node[right=of baselines, yshift=-0.2cm, xshift=-0.5cm](DIP){\includegraphics[width=7.8cm]{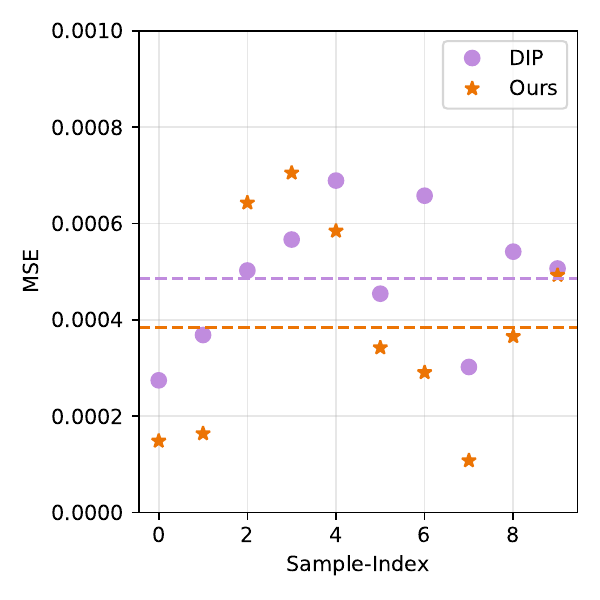}};
        \node[above=of DIP, yshift=-2.5cm, xshift=-0.75cm]{$\bar{t}_\text{DIP}\approx$ 6915 s};
        \node[above=of DIP, yshift=-1.5cm, xshift=-3.75cm]{(b)};
    \end{tikzpicture}
    }
    \vspace{-0.5cm}
    \caption{Evaluation of Experiment B in projection domain. (a) MSE distribution of 300 test data with respect to measured raw data for different reconstruction methods. The mean reconstruction time $\bar{t}$ is denoted for each method. (b) Comparison of \methodname\ and the DIP baseline paired per-slice MSE points for 10 slices. Mean values are indicated by horizontal lines. Values in the projection domain are unit-less.}
    \label{fig:B-proj}
\end{figure}

In Experiment B, real projection data was used for reconstruction. The given image slices from the 2DeteCT dataset using FBP merely pose one possible reconstruction result. Thus, there exists no true ground-truth image information in Experiment B and the comparison to the supervised approach as well as the quantitative MSE/SSIM evaluation in the image domain should be omitted. 
Again, the projection domain difference to the measured data was evaluated. As no GT was available, the supervised network could not be applied. 
Figure~\ref{fig:B-proj}(a) shows the distribution of MSE of the test datasets for all considered methods.
It can be observed that, in terms of similarity to raw data, FBP and ML perform slightly worse than the network reconstruction. The deviations of this figure also characterize the forward-model mismatch between forward projected image results and real measurement data.

\begin{figure}[]
    \centering
    \resizebox{12cm}{!}{
    \begin{tikzpicture}
        \node[](img){\includegraphics[width=11cm, trim=0cm 20cm 0cm 0cm, clip]{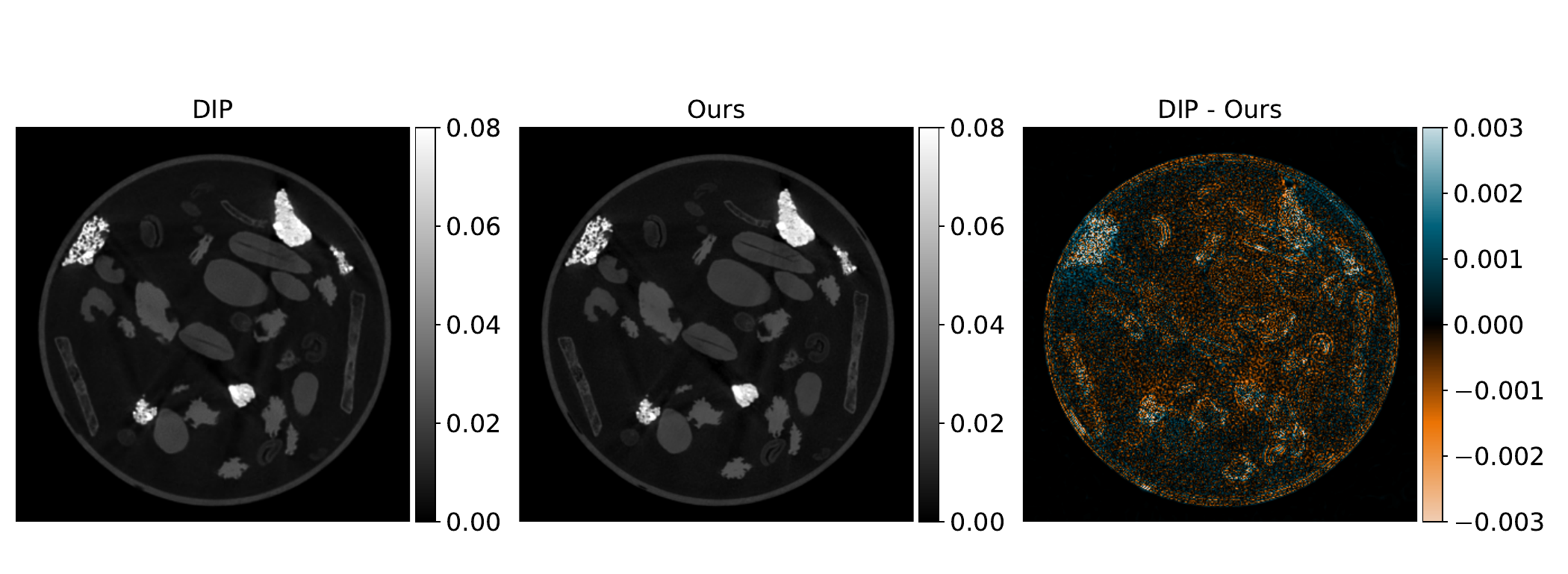}};
        \node[below=of img, yshift=+1.5cm](img2){\includegraphics[width=11cm, trim=0cm 12cm 0cm 15cm, clip]{images/eval_dip_vs_ours_90.pdf}};
        \node[below=of img2, yshift=+1.5cm]{\includegraphics[width=11cm, trim=0cm 0cm 0cm 24cm, clip]{images/eval_dip_vs_ours_90.pdf}};
    \end{tikzpicture}
    }
    \vspace{-0.75cm}
    \caption{Comparison of \methodname\ and the DIP baseline on an exemplary qualitative comparison in image and projection domain. Image values are linear attenuation coefficients in $1/\text{cm}$. 
    }
    \label{fig:dip}
\end{figure}

We compare our results against the DIP baseline on 10 selected slices. Figure~\ref{fig:B-proj}(b) presents paired data points for projection domain MSE. In terms of reconstruction quality, DIP and our method are comparable (MSE DIP $\approx 5\times10^{-4}$, MSE \methodname $\,\approx 4\times10^{-4}$ unitless). This can also be observed for the qualitative example shown in Figure~\ref{fig:dip}. 

It should be noted that the reconstruction for our method only requires pure network inference. For 10 reconstructions, we observe a mean reconstruction time of 438 ms ($\pm$ 51 ms) per dataset for the network, whereas FBP takes 1717 ms ($\pm$ 47 ms) and ML takes approx. 63\,400 ms ($\pm$ 2\,200 ms) on average. On full resolution of $1024 \times 1024$ pixels, DIP requires a per-image optimization runtime of 6915~s ($\approx$ 115~min) $\pm$ 109~s- This means our method facilitates a speed-up of 15000$\times$ for a comparable reconstruction quality.

\begin{figure}[t!]
    \centering
    \sffamily
    \resizebox{0.95\textwidth}{!}{%
    \begin{tikzpicture}
    \node[inner sep=0pt] (img) {\includegraphics[width=17.6cm]{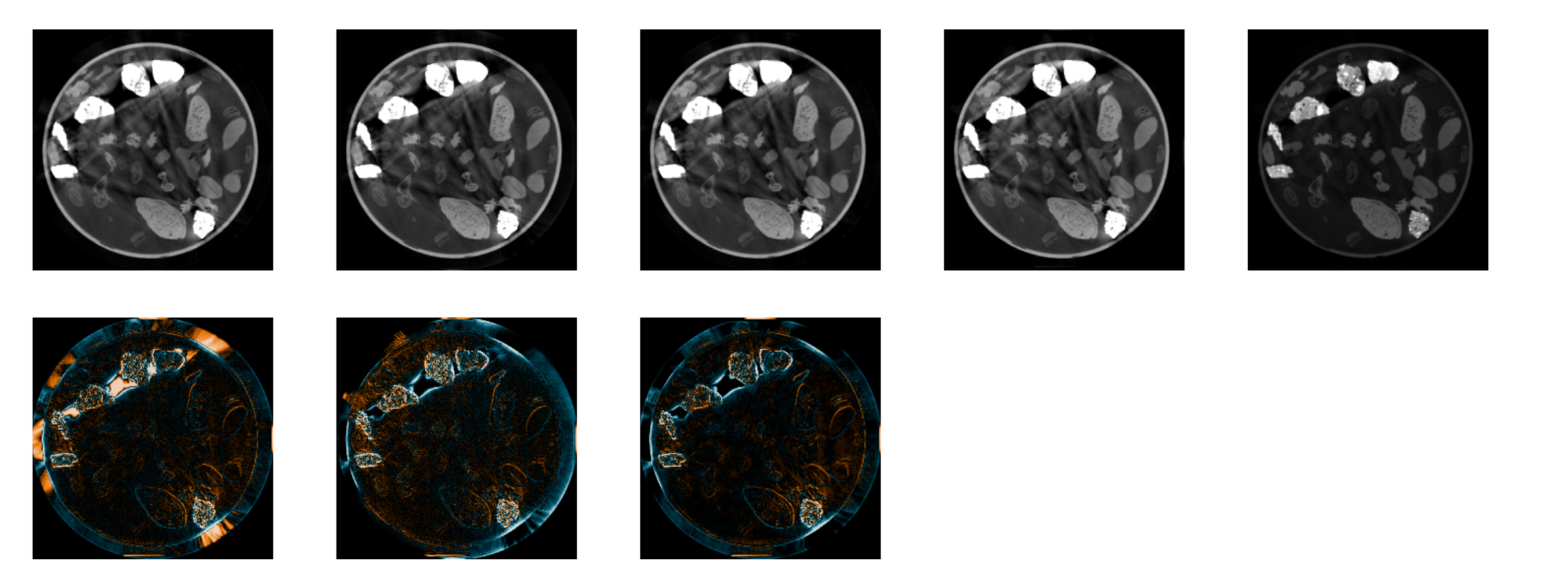}};
    \node at ([yshift=-0.25cm, xshift=-0.1cm]img.north west) {(a)};
    \node at ([yshift=-0.145cm, xshift=1.75cm]img.north west) {\small FBP};
    \node at ([yshift=-0.145cm, xshift=5.15cm]img.north west) {\small ML};
    \node at ([yshift=-0.155cm, xshift=8.55cm]img.north west) {\small Supervised};
    \node at ([yshift=-0.145cm, xshift=11.95cm]img.north west) {\small \methodname};
    \node at ([yshift=-0.145cm, xshift=15.35cm]img.north west) {\small GT$^\star$};
    \node at ([yshift=-3.4cm, xshift=1.75cm]img.north west) {\small FBP - \methodname};
    \node at ([yshift=-3.4cm, xshift=5.15cm]img.north west) {\small ML - \methodname};
    \node at ([yshift=-3.41cm, xshift=8.55cm]img.north west) {\small Supervised - \methodname};
    \draw[orange, thick] (5.45,1.6) rectangle (6.55,2.7);
    \node at (2.2,-1.6) {
        \begin{axis}[
            anchor=north west, colorbar,
            colormap={custom_shift}{
            rgb255(0cm)=(240,205,179); rgb255(1cm)=(236,116,4);
            rgb255(2cm)=(0,0,0); rgb255(3cm)=(0,97,122);
            rgb255(4cm)=(198,220,226)},
            colorbar style={ width=0.5cm, height=2.7cm, ytick={0,1,2,3,4},
                yticklabels={$-0.006 \frac{1}{\mathrm{cm}}$, $+0.006 \frac{1}{\mathrm{cm}}$},},
            axis y line=none, axis x line=none, xmin=0, xmax=1, ymin=0, ymax=1,
        ]\end{axis}
    };
    \end{tikzpicture}
    }
    \resizebox{0.95\textwidth}{!}{%
    \begin{tikzpicture}
    \node[inner sep=0pt] (img) {\includegraphics[width=17.6cm]{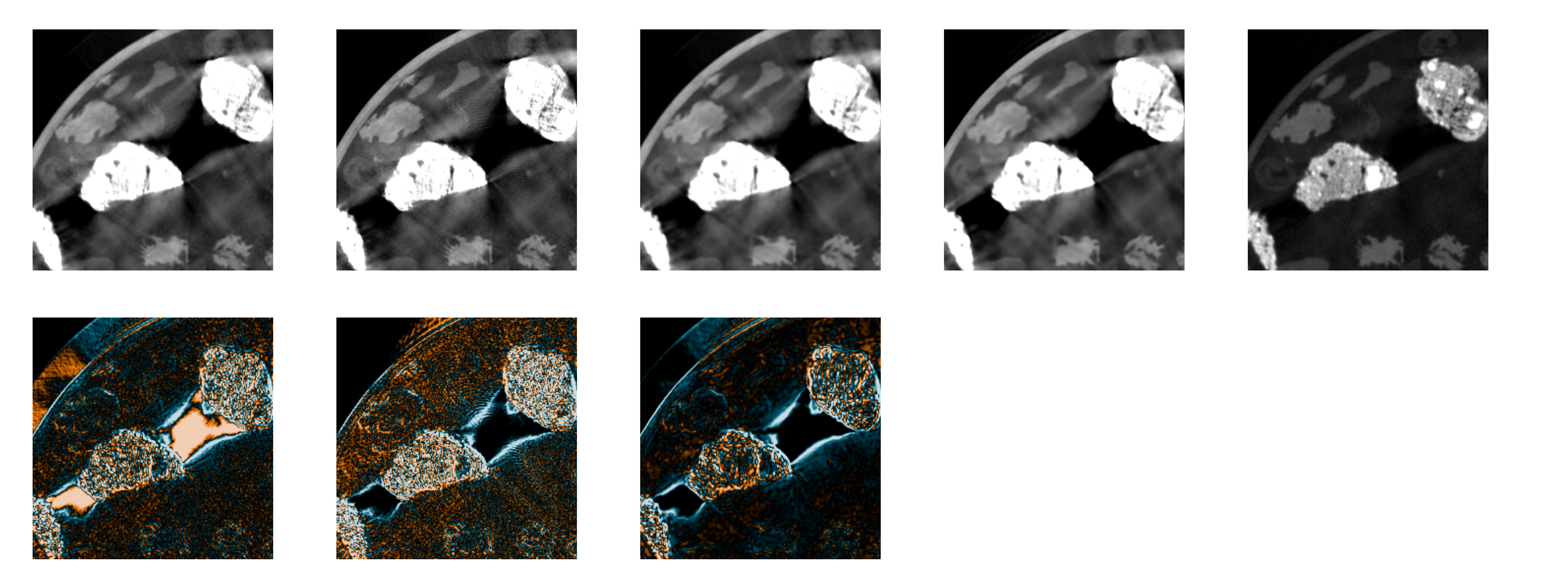}};
    \node at ([yshift=-0.25cm, xshift=-0.1cm]img.north west) {(b)};
    \node at ([yshift=-0.145cm, xshift=1.75cm]img.north west) {\small FBP};
    \node at ([yshift=-0.145cm, xshift=5.15cm]img.north west) {\small ML};
    \node at ([yshift=-0.155cm, xshift=8.55cm]img.north west) {\small Supervised};
    \node at ([yshift=-0.145cm, xshift=11.95cm]img.north west) {\small \methodname};
    \node at ([yshift=-0.145cm, xshift=15.35cm]img.north west) {\small GT$^\star$};
    \node at ([yshift=-3.4cm, xshift=1.75cm]img.north west) {\small FBP - \methodname};
    \node at ([yshift=-3.4cm, xshift=5.15cm]img.north west) {\small ML - \methodname};
    \node at ([yshift=-3.41cm, xshift=8.55cm]img.north west) {\small Supervised - \methodname};
    \draw[orange, thick] (5.19,0.27) rectangle (7.91,2.98);
    \draw[orange, thick] (5.17,0.25) rectangle (7.93,3.00);
    \draw[-{Latex}, thick, color=orange] (-4.3,0.8) -- ($(-4.3,0.8)+(+0.4,-0.2)$);
    \draw[-{Latex}, thick, color=orange] (-0.9,0.8) -- ($(-0.9,0.8)+(+0.4,-0.2)$);
    \draw[-{Latex}, thick, color=orange] (2.5,0.8) -- ($(2.5,0.8)+(+0.4,-0.2)$);
    \draw[-{Latex}, thick, color=orange] (5.9,0.8) -- ($(5.9,0.8)+(+0.4,-0.2)$);
    \node at (2.2,-1.6) {
        \begin{axis}[
            anchor=north west, colorbar,
            colormap={custom_shift}{
            rgb255(0cm)=(240,205,179); rgb255(1cm)=(236,116,4);
            rgb255(2cm)=(0,0,0); rgb255(3cm)=(0,97,122);
            rgb255(4cm)=(198,220,226)},
            colorbar style={ width=0.5cm, height=2.7cm, ytick={0,1,2,3,4},
                yticklabels={$-0.006 \frac{1}{\mathrm{cm}}$, $+0.006 \frac{1}{\mathrm{cm}}$},},
            axis y line=none, axis x line=none, xmin=0, xmax=1, ymin=0, ymax=1,
        ]\end{axis}
    };
    \end{tikzpicture}
    }
    \caption{Reconstructed images for different methods in level/window=0.03/0.06\(\,\frac{1}{\mathrm{cm}}\). GT$^\star$ is generated from high-quality acquisition mode projection data without beam-hardening and is a high-quality FBP reconstruction, \emph{not} a true ground truth. Difference images are shown with respect to our method.}
    \label{fig:B-qualitative}
\end{figure}

Figure~\ref{fig:B-qualitative} shows the qualitative comparison in the image domain. The deviations seem to lie mostly in edges, higher density as well as artifact regions.
The orange arrows in Figure~\ref{fig:B-qualitative}(b) indicate that some details might get lost for the supervised and our approach due to smoothing.


\subsubsection{Ablation studies}
\begin{figure}[ht!]
\centering
\includegraphics[width=5.5cm]{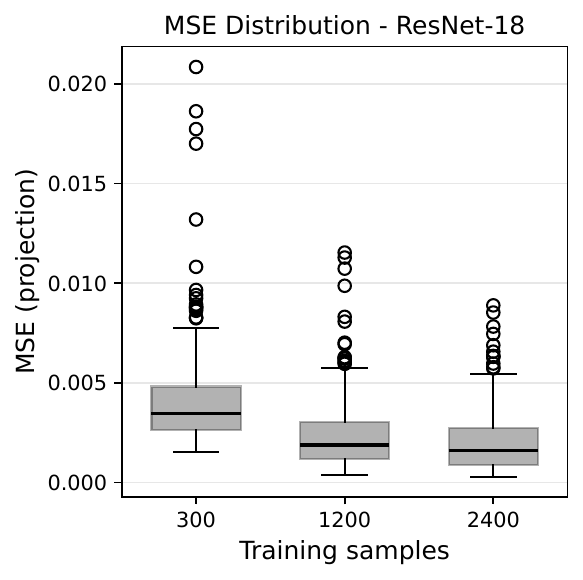}
\includegraphics[width=5.5cm]{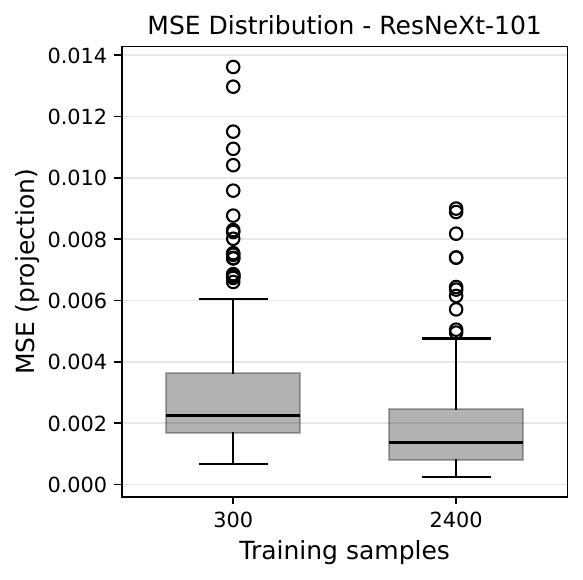}
\caption{Data-efficiency of the proposed training scheme for different network architectures: Reconstruction metrics on test data as a function of training-set size.}
\label{fig:datasize}
\end{figure}

The dataset-size sweep (300/1200/2400 training slices) indicates a monotonic improvement in reconstruction quality with more training data. The corresponding trend is shown in Figure~\ref{fig:datasize}. The architecture-swap with lower resolution confirms interchangeability of the backbone: Replacing the ResNeXt-101 encoder with a smaller ResNet-18 backbone yields higher MSE values while preserving the overall behavior.

\subsubsection{Stability across seeds}
Repeating the training over three random seeds (42, 123, 456) for a maximum of 500 epochs, using the ResNet-18 backbone with 300 training samples at a resolution of $256\times256$ pixels, yields the results shown in Table~\ref{tab:training_time}. To assess seed-dependent variability, we computed the standard deviation across the three seeds for each sample, normalized by the data range (image domain) or the global mean MSE (projection domain). In the training projection data, the resulting variability of $\approx 14\%$ reflects differing states of convergence.
Stable behavior would require this spread to be comparable to that observed on the test data. Since the relative deviation of the reconstructed image pixels on the test set amounts to only $\approx 8\%$, this criterion is satisfied.
\begin{table}[]
  \centering
  \caption{Comparison across seeds for 300 training slices, ResNet-18 backbone, and max. number of 500 epochs}
  \label{tab:training_time}
  \begin{tabular}{lcccc}
    \hline
    \textbf{Metric} & \textbf{Seed 1} & \textbf{Seed 2} & \textbf{Seed 3} &  \\
    \hline
    Training Time & 28.9 h & 27.0 h & 27.7 h \\    
    Best Epoch & 455 & 496 & 494  \\
    MSE, Proj., Train & 3.7 $\times 10^{-3}$  & 4.1 $\times 10^{-3}$ & 3.5 $\times 10^{-3}$   \\
    MSE, Proj. Test & 4.1 $\times 10^{-3}$ & 5.2 $\times 10^{-3}$  & 4.4 $\times 10^{-3}$  \\
    \hline 
    & & \textbf{Across Seeds} & \\
    MSE Variability, Proj., Train & & 14.7\%  $\pm$ 6.9\%\\
    Pixel Variability, Image, Test & &  8.3\% $\pm$ 5.3\% \\
    \hline
  \end{tabular}
\end{table}


\subsubsection{Noise robustness}
\begin{figure}[]
\sffamily
\centering
\resizebox{\textwidth}{!}{
\begin{tikzpicture}
    \node[](img){\includegraphics[width=14cm]{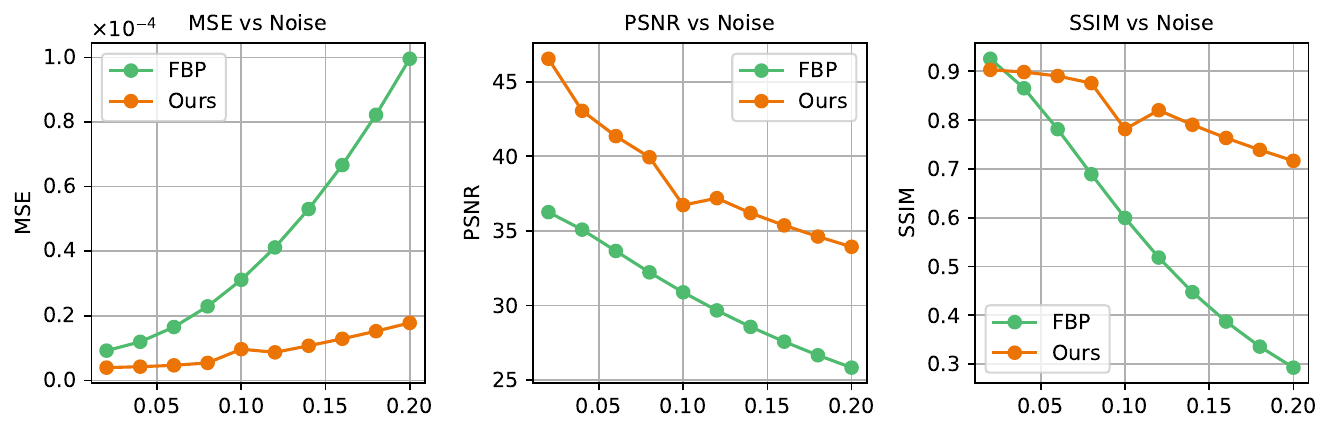}};
    \node[below=of img, xshift=+0.5cm, yshift=1.35cm]{\footnotesize Noise level $\sigma$};    
    \node[below=of img, xshift=+5.2cm, yshift=1.35cm]{\footnotesize Noise level $\sigma$};
    \node[below=of img, xshift=-4.2cm, yshift=1.35cm]{\footnotesize Noise level $\sigma$};
\end{tikzpicture}
}
\vspace{-0.5cm}
\caption{Noise-robustness of FBP vs \methodname\ at fixed geometry across different noise levels. MSE in $1/\text{cm}^2$. PSNR denotes peak signal to noise ratio.}
\label{fig:noise}
\end{figure}

To assess robustness to measurement noise at fixed geometry, we vary the noise level by adding Gaussian noise in the projection domain (after log-transform) and evaluate reconstruction metrics.
Across the tested noise levels of approximately $0.25\%$ to $5\%$ of maximal projection value ($\sigma \in[0.01,0.2]$), reconstruction metrics w.r.t. clean reconstruction degrade gracefully with increasing noise as shown in comparison with FBP reconstruction in Figure~\ref{fig:noise}.

\section{Discussion}\label{sec:discussion}

\noindent \textbf{Does our approach perform better than FBP and ML in Experiment A?} Yes, within this controlled, geometry-matched benchmark. The reference image is the mathematically optimal solution for the pure Radon inversion task. This is a controlled environment with perfect forward-model knowledge, and no broad clinical superiority should be inferred from it.
As observed in Figure~\ref{fig:A-quantitative} and Figure~\ref{fig:A-qualitative} the network predicts reconstructed images closer to the reference than FBP and ML. The supervised network performs similarly with slightly better results.

For high-quality data (Experiment A), FBP outperforms ML in the default ML configuration. From the NPS analysis (Figure~\ref{fig:A-NPSandMTF}), we attribute this result to the increased noise distribution in the images. As can be observed in the real-data case (Experiment B), ML does provide better results with the same tolerance choices.
The noise analysis shows that the deep learning methods tend to smooth images more than FBP and ML.  
We discuss the clinical implications of this over-smoothing honestly. As quantified by the MTF and NPS analyses, the noise reduction does not strongly compromise the measured spatial resolution. Nevertheless, pixel- and frequency-based metrics cannot fully guarantee the preservation of small, low-contrast, diagnostically relevant structures. A task-based detectability study is left to future work, and we frame the present results as a feasibility study rather than a diagnostic validation.\\


\noindent \textbf{How does the approach perform in Experiment B?}
Figure~\ref{fig:B-proj} shows that adherence to the raw projection datafor our reconstruction network is better than FBP, and comparable to ML in the given setting.
When inspecting images in Figure~\ref{fig:B-qualitative}, some details seem lost for both the supervised and the proposed unsupervised method. We hypothesize this stems from the observed tendency towards smooth reconstructions (as was discussed for Experiment A). In addition, the forward-model mismatch between the linear Radon model and the beam-hardening physics is a plausible contributing cause, consistent with the residual analysis reported above.
The DIP baseline confirms that a projection-domain loss combined with a convolutional architecture can reconstruct real data without image-domain ground truth, reaching a reconstruction quality comparable to our approach (Fig.~\ref{fig:dip}). The decisive difference lies in inference: DIP re-optimizes the network weights for every individual scan, whereas our amortized framework reconstructs any unseen scan in a single forward pass, a speed-up of about $15000\times$. This illustrates that the central benefit of our method is not a fundamentally different image prior, but the transfer from per-instance optimization to amortized single-pass inference across a dataset.

\subsection*{Limitations}\label{sec:limitations}
The proposed approach presents several notable limitations.
Firstly, this work is confined to two-dimensional reconstructions on a single dataset with fixed geometry. 
Since the discretized Radon transform is ill-conditioned for a limited number of projections, many images yield near-identical projection data, and the data term alone cannot determine a unique solution. Instead of an explicit penalty, our framework relies on the implicit prior induced by the network architecture, whose convolutional, encoder–decoder structure biases the solution towards spatially coherent, natural-image-like reconstructions. In this sense the network acts as a learned image-domain filter, replacing the hand-crafted regularizer of classical variational approaches~\cite{Ulyanov2017, Dittmer2019}.
The study uses only the 2DeteCT dataset, without multi-dataset validation, and we do not claim direct transferability without re-training or fine-tuning for new datasets.
Furthermore, no data augmentations were applied, which preserves geometric consistency but may reduce network generalization to other field-of-views of the input image. Applicable augmentations were discussed previously~\cite{Hellwege2026Augmentations} but were not considered in this work.
The method is furthermore sensitive to forward-model mismatch, as seen in Experiment~B (linear Radon versus beam hardening), and it exhibits a tendency towards over-smoothing that may affect the detectability of small structures.
While the training process is time-consuming, the application of the trained network is executed with notable speed. \\


\subsection*{Future work}\label{sec:future}
Based on the observed limitations, we propose the following advancements.
To expand to 3D reconstruction, the approach should be tested with cone-beam CT datasets.
We assume that the observed smoothing results from the pixel-wise projection-domain loss combined with the low-frequency bias of the convolutional architecture. Future work could counteract this using a perceptual loss~\cite{Johnson2016}, or adversarial training~\cite{Wolterink2017} to restore high-frequency texture. Both are compatible with our framework and require only a set of representative images rather than paired ground-truth samples.
The geometry of the projection layers should be made adaptable to the dataset. A concrete roadmap includes: Meta-learning for geometry generalization, training and validation on multiple datasets, and more advanced uncertainty quantification e.g. through ensemble methods. Additionally, a task-based detectability study will improve clinical impact.
Overall, the inclusion of unsupervised neural networks into image reconstruction creates new opportunities for use-case adaptation.
While we focused on CT imaging, the general approach can be transferred to arbitrary inverse problems for which a continuous forward model (Eq.~\ref{eq:inverseProblem}) can be modeled.

\section*{Data statement}
All data analyzed in this study are included in the 2DeteCT dataset~\cite{Kiss2023}. A minimal working example is available at our repository \href{https://gitlab.com/maik.stille/unsuperviseddlreco}{https://gitlab.com/maik.stille/unsuperviseddlreco}.

\section*{Author Contribution Statement}
L.H. wrote the main manuscript text. L.H. and J.C.E. prepared figures and conceptualized the experiment setup. M.Sch. supported experiment design and evaluation. M.St. offered critical reviews and provided essential guidance in defining the research question. All authors critically reviewed the manuscript. Final proofreading was performed by T.M.B..

\section*{Author Statement}
The authors declare no competing interests.

\section*{Ethics Declaration}
Not applicable.

\section*{Generative AI Statement}
During the preparation of this work, the authors used Claude Opus 4.8/Anthropic in order to improve structure, language and readability of the manuscript text. After using this tool, the authors reviewed and edited the content as needed and take full responsibility for the content of the publication.

\section*{Funding}
This work was partially funded by the state of Schleswig-Holstein through the project "Individualisierte Medizintechnik für bildgestützte, robotische Interventionen (IMTE 2)", project number: 125 24 009.

\printbibliography

\end{document}

%% file: gradient_based_reco_scheme.tex
\sffamily
    \centering
    \resizebox{0.75\textwidth}{!}{%
    \begin{tikzpicture}[
        boxblack/.style={draw=none, fill=black, text width=1.7cm, text centered, minimum height=2.5cm, minimum width=2.5cm,font=\huge},
        boxorange/.style={draw=none, fill=orange!20!, text width=1.7cm, text centered, minimum height=2.5cm, minimum width=2.5cm, font=\huge},
        boxblue/.style={draw=none, fill=mylightblue, text width=1.7cm, text centered, minimum height=2.5cm, minimum width=2.5cm,font=\huge},
        boxgrey/.style={draw=none, fill=white!90!gray, text width=1cm, text centered, minimum height=4cm, font=\huge},
        ellipseboxorange/.style={draw=orange, circle, minimum height=1cm, minimum width=2.5cm, fill=orange!10, font=\huge},
        rect/.style={draw, rectangle, minimum height=1cm, minimum width=2.5cm, fill=orange!20, text centered, font=\huge},
        arroworange/.style={->, line width=1mm, orange, >=stealth},
        arrowgray/.style={->, line width=1mm, mylightblue!75!black, >=stealth},
        arrowgreen/.style={->, line width=1mm, mygreen!75!black, >=stealth},
        arrowgray/.style={->, line width=1mm, gray, >=stealth},
        dashedarroworange/.style={->, line width=1mm, orange, dashed, >=stealth},
        dashedarrowgray/.style={->, line width=1mm, mylightblue!75!black, dashed, >=stealth}, 
        node distance=1cm
    ]

    \node[boxblue, minimum width=2.8cm, minimum height=2.8cm, anchor=center] (image0) {};
    \node[draw=none, fill=black!80!white, text width=1.7cm, text centered, minimum height=2.5cm, minimum width=2.5cm,font=\huge] {};
    \node[draw=none, above=of image0, font=\large, align=center, anchor=center] (labelBP) {Start: \\ Initial image $\hat{f}^0$};

    \node[boxorange, minimum width=2.8cm, minimum height=2.8cm, below=of image0] (y) {};
    \node[anchor=center] at (y) {\includegraphics[height=2.5cm, width=2.5cm, angle=90] {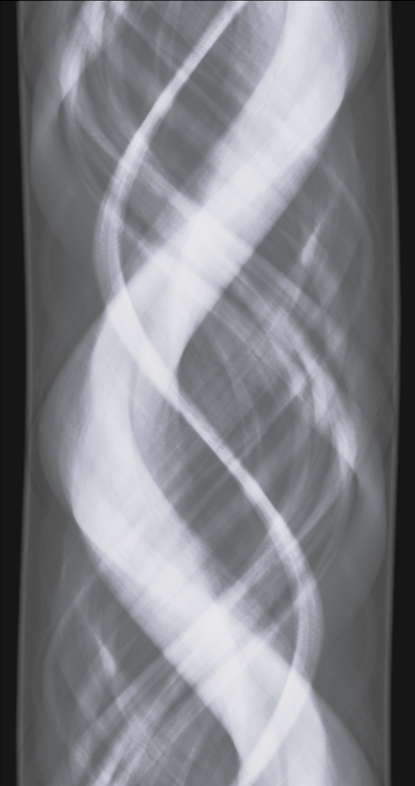}};
    \node[draw=none, above=of y, yshift=-0.5cm, font=\large, anchor=center, anchor=center] (labelMeasurement){Measurements $p$};
    
    \node[boxblue, right=of image0, xshift=2.5cm, minimum width=2.8cm, minimum height=2.8cm, anchor=center] (image) {};
    \node[anchor=center] at (image) {\includegraphics[width=2.5cm] {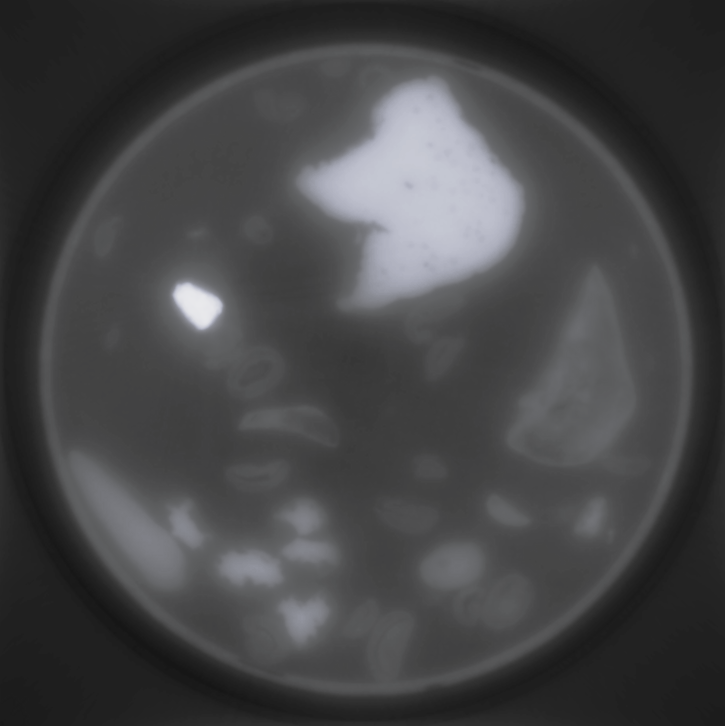}};
    \node[draw=none, above=of image, font=\large, align=center, anchor=center] (labelPred) {Current \\ image $\hat{f}^k$};

    \draw[dashedarrowgray] (image0) -- (image);
    \node[draw=none, right=of image0, xshift=-0.7cm, yshift=+0.5cm] (FP) {\large $k=0$};

    \node[boxorange, right=of image, minimum width=2.8cm, minimum height=2.8cm, xshift=2.5cm, anchor=center] (proj) {};
    \node[anchor=center] at (proj){\includegraphics[height=2.5cm, width=2.5cm, angle=90]{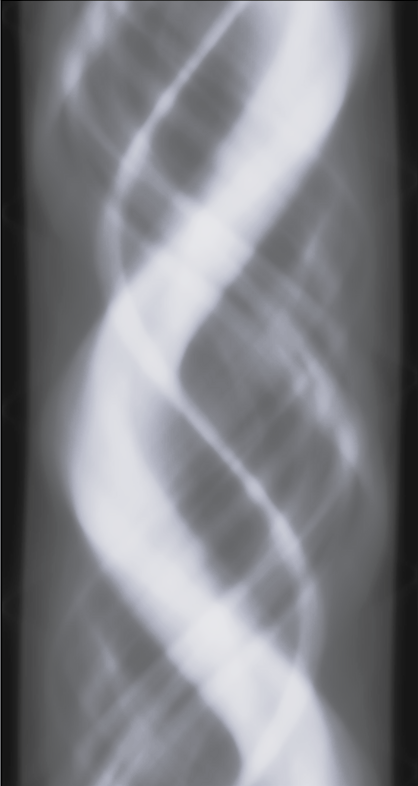}};
    \node[draw=none, above=of proj, font=\large, align=center, anchor=center] {Projections $p^k$ of\\ current image };

    \node[boxgrey, text width=1.1cm, minimum height=3cm, right=of image, xshift=-0.5cm, yshift=0cm] {};
    \node[draw=none, right=of image, xshift=-0.3cm, yshift=+0.7cm] (FP) {\textcolor{gray!70!black}{\Large $A$}};
    \draw[arrowgray] (image.10) -- (proj.170);
    \draw[dashedarroworange] (proj.190) -- (image.-10);
    \node[draw=none, right=of image, xshift=-0.3cm, yshift=-0.7cm] (FP) {\textcolor{orange!90!black}{\Large $A^T$}};

    \node[ellipseboxorange, below=of image, yshift=-1.4cm, font=\Large, anchor=center] (L) {$\mathcal{L}$};
    \node[draw=none, right=of L, xshift=-1cm, yshift=0.75cm, font=\huge] {\textcolor{orange}{$\nabla$}};

    \draw[dashedarrowgray] (y) -- (L);
    \draw[dashedarroworange, yshift=-1cm] (L.10) -| (proj.-105);
    \draw[dashedarrowgray] (proj.-85) |- (L.-10);

    

    \draw[arrowgray] (11.0, -3.0) -- (11.8, -3.0);
    \node[draw=none, rectangle, anchor=west, font=\large] at (12.0, -3.0) {\textcolor{gray}{Forward pass}};
    \draw[dashedarrowgray] (11.0, -3.5) -- (11.8, -3.5);
    \node[draw=none, rectangle, anchor=west, font=\large] at (12.0, -3.5) {\textcolor{gray}{Loss calculation}};
    \draw[dashedarroworange] (11.0, -4.0) -- (11.8, -4.0);
    \node[draw=none, rectangle, anchor=west, font=\large] at (12.0, -4.0) {\textcolor{orange!90!black}{Gradient step}};

    \end{tikzpicture}
    }

%% file: training_scheme.tex
\sffamily
    \centering
    \resizebox{0.9\textwidth}{!}{%
    \begin{tikzpicture}[
        boxorange/.style={draw=none, fill=orange!20!, text width=1.7cm, text centered, minimum height=2cm, font=\huge},
        boxblue/.style={draw=none, fill=mylightblue, text width=1.7cm, text centered, minimum height=2cm, font=\huge},
        boxgrey/.style={draw=none, fill=gray!10!white, text width=1cm, text centered, minimum height=4cm, font=\huge},
        ellipseboxorange/.style={draw=orange, circle, minimum height=1cm, minimum width=2cm, fill=orange!10, font=\huge},
        rect/.style={draw, rectangle, minimum height=1cm, minimum width=2cm, fill=orange!20, text centered, font=\huge},
        arroworange/.style={->, line width=1mm, orange, >=stealth},
        arrowgray/.style={->, line width=1mm, mylightblue!75!black, >=stealth},
        arrowgreen/.style={->, line width=1mm, mygreen!75!black, >=stealth},
        arrowgray/.style={->, line width=1mm, gray, >=stealth},
        dashedarroworange/.style={->, line width=1mm, orange, dashed, >=stealth},
        dashedarrowgray/.style={->, line width=1mm, mylightblue!75!black, dashed, >=stealth}, 
        node distance=1cm
    ]
    
    \node[boxorange, minimum width=2.8cm, minimum height=2.8cm] (measurements) {};
    \node[anchor=center] at (measurements) {\includegraphics[height=2.5cm, width=2.5cm] {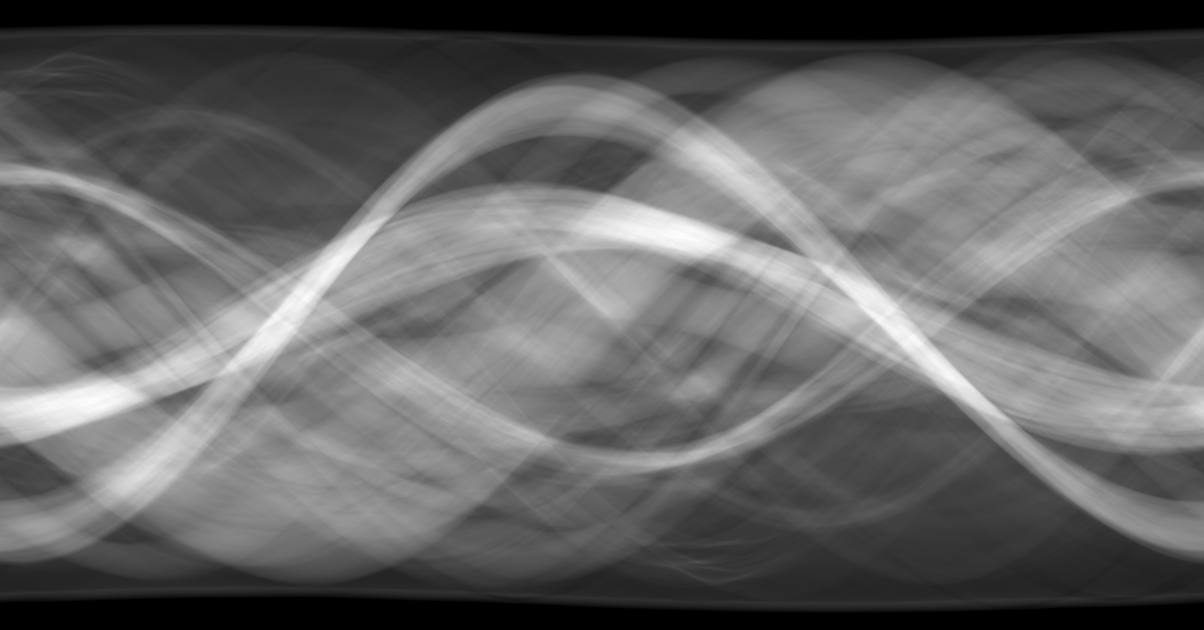}};
    \node[draw=none, above=of measurements, yshift=-0.5cm, font=\large, align=center] (labelMeasurement){Measurements $p$};
    
    \node[boxgrey, text width=1.1cm, minimum height=3cm, right=of measurements, xshift=-0.75cm, yshift=0cm] (boxBP) {};
    \node[draw=none, anchor=center, yshift=+0.5cm] at (boxBP) {\textcolor{gray!75!black}{\Large $A^T$}};

    \node[boxblue, right=of measurements, xshift=2.5cm, minimum width=2.8cm, minimum height=2.8cm, anchor=center] (imageBP) {};
    \node[anchor=center] at (imageBP) {\includegraphics[width=2.5cm] {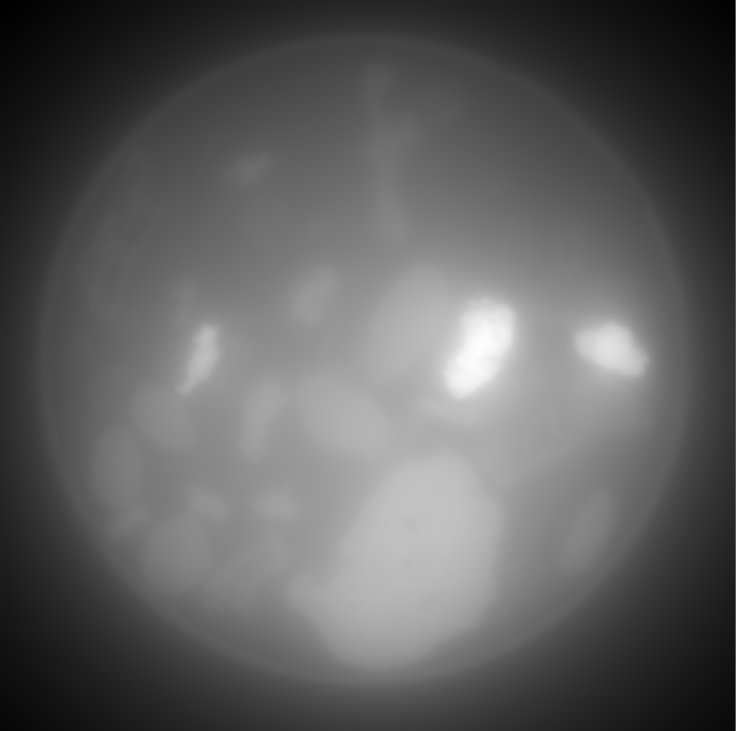}};
    \node[draw=none, above=of imageBP, yshift=-0.5cm, font=\large, align=center] (labelBP) {Image $\tilde{f}$\\ from BP};
        
    \draw[arrowgray] (measurements) -- (imageBP);

    \node[draw=black, circle, right=of imageBP, font=\huge] (Unet) {\includegraphics[width=1.2cm]{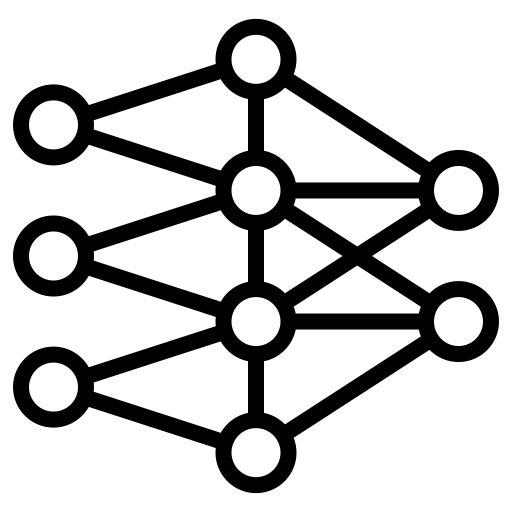}};
    \node[draw=none, minimum size=0.1cm, above=of Unet, yshift=-0.7cm, font=\large] (labelUnet) {UNet++};
    
    \draw[arrowgray] (imageBP) -- (Unet);
    
    \node[boxblue, right=of Unet, xshift=1.5cm, minimum width=2.8cm, minimum height=2.8cm, anchor=center] (image) {};
    \node[anchor=center] at (image) {\includegraphics[width=2.5cm] {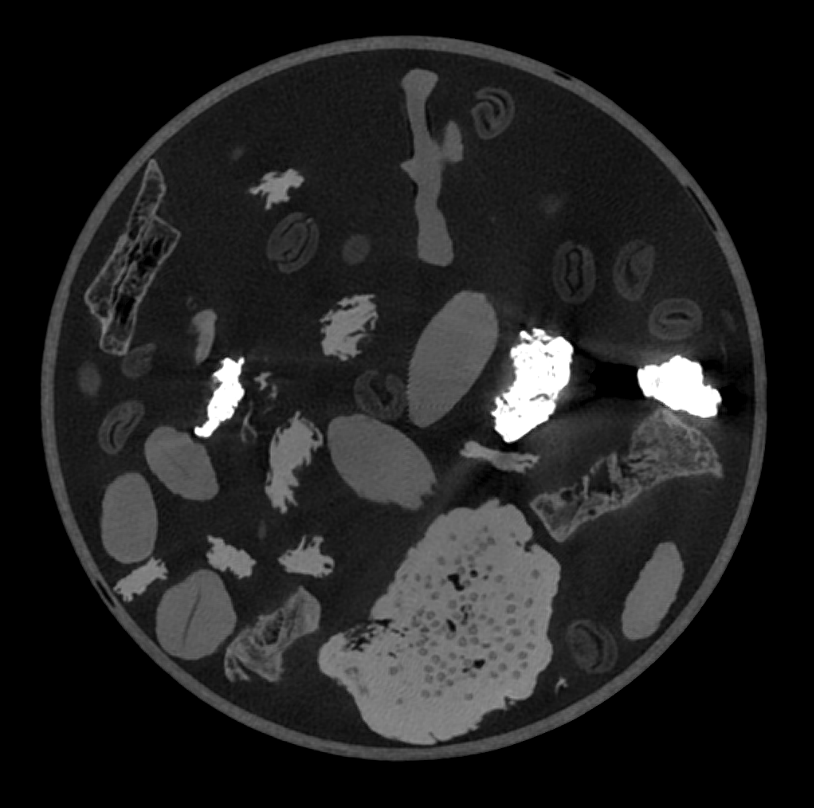}};
    \node[draw=none, above=of image, font=\large, align=center, anchor=center] (labelPred) {Prediction $\hat{f}$};
    
    \draw[arrowgray] (Unet.10) -- (image.173);
    \draw[dashedarroworange] (image.187) -- (Unet.-10);

    \node[boxorange, right=of image, minimum width=2.8cm, minimum height=2.8cm, xshift=2.5cm, anchor=center] (proj) {};
    \node[anchor=center] at (proj){\includegraphics[height=2.5cm, width=2.5cm]{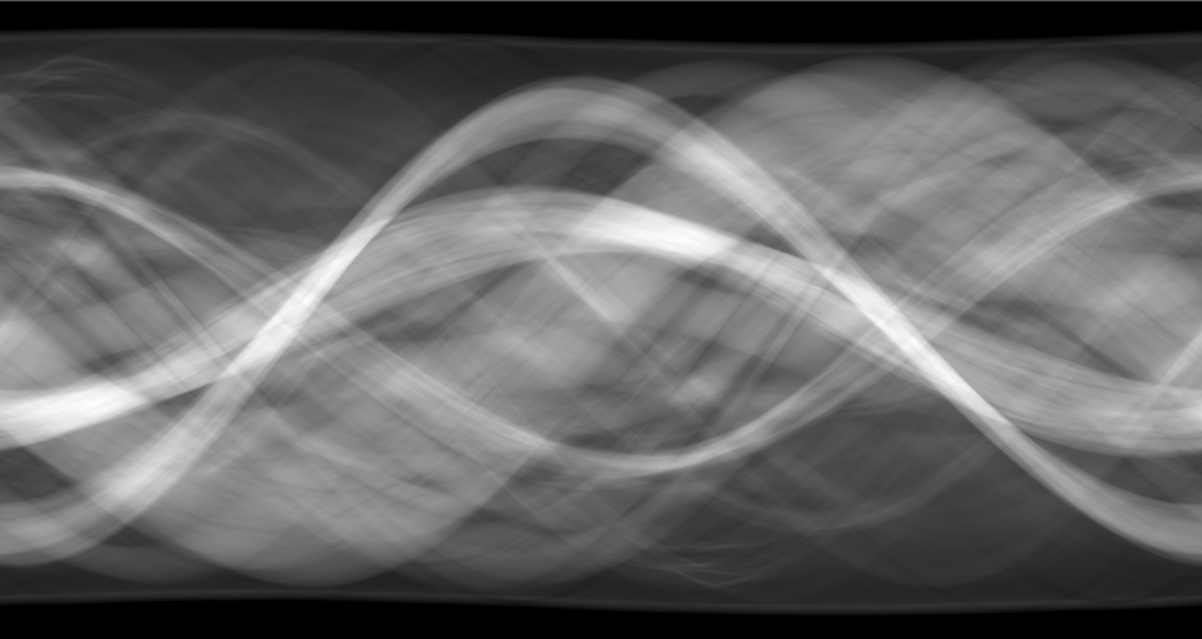}};
    \node[draw=none, above=of proj, font=\large, align=center, anchor=center] {Projections $p^k$ of\\ current image };

    \node[boxgrey, text width=1.1cm, minimum height=3cm, right=of image, xshift=-0.5cm, yshift=0cm] {};
    \node[draw=none, right=of image, xshift=-0.3cm, yshift=+0.7cm] (FP) {\textcolor{gray!75!black}{\Large $A$}};
    \draw[arrowgray] (image.10) -- (proj.170);
    \draw[dashedarroworange] (proj.190) -- (image.-10);
    \node[draw=none, right=of image, xshift=-0.3cm, yshift=-0.7cm] (FP) {\textcolor{orange!90!black}{\Large $A^T$}};

    \node[ellipseboxorange, below=of Unet, yshift=0.5cm] (L2) {\Large $\mathcal{L}$};
    \node[draw=none, right=of L2, xshift=-0.5cm, yshift=0.75cm, font=\huge] {\textcolor{orange}{$\nabla$}};

    \draw[dashedarrowgray] (measurements) |- (L2);
    \draw[dashedarroworange, yshift=-1cm] (L2.10) -| (proj.-105);
    \draw[dashedarrowgray] (proj.-85) |- (L2.-10);

    

    
    \draw[arrowgray] (13.0, -3.5) -- (13.7, -3.5);
    \node[draw=none, rectangle, align=left] at (15.0, -3.5) {\textcolor{gray}{Forward pass}};
    \draw[dashedarrowgray] (13.0, -4.0) -- (13.7, -4.0);
    \node[draw=none, rectangle, align=left] at (15.2, -4.0) {\textcolor{gray}{Loss calculation}};
    \draw[dashedarroworange] (13.0, -4.5) -- (13.7, -4.5);
    \node[draw=none, rectangle, align=left] at (15.1, -4.5) {\textcolor{orange}{Backward pass}};    

    
    \end{tikzpicture}
    }